\documentclass[amsmath,amssymb,superscriptaddress,aps,prl,twocolumn,showpacs,reprint]{revtex4-1}
\pdfoutput=1

\usepackage{amsmath}
\usepackage{bm}
\usepackage{color}
\usepackage{graphicx}
\usepackage{multirow}

\renewcommand {\vec} [1] {{\bm #1}}

\begin{document}

\title{Prediction of the derivative discontinuity in density
  functional theory from an electrostatic description of the exchange
  and correlation potential}

\author{Xavier Andrade} 
\email{xavier@tddft.org}
\author{Al\'an Aspuru-Guzik}
\email{alan@aspuru.com}
\affiliation{Department of Chemistry and Chemical Biology, Harvard
  University, 12 Oxford Street, Cambridge, MA 02138, United States}


\begin{abstract}
  We propose a new approach to approximate the exchange and
  correlation (XC) functional in density functional theory. The XC
  potential is considered as an electrostatic potential, generated by
  a fictitious XC density, which is in turn a functional of the
  electronic density. We apply the approach to develop a correction
  scheme that fixes the asymptotic behavior of any approximated XC
  potential for finite systems. Additionally, the correction procedure
  gives the value of the derivative discontinuity;
  therefore it can directly predict the fundamental gap as
  a ground-state property.
\end{abstract}

\maketitle

An important and long standing topic in density functional theory
(DFT)~\cite{Hoh1964PR,*Koh1965PR} is the prediction of the fundamental
gap~\cite{Coh2009Sci,*Per1985IJQC,Gru2006JCP} \(E_\mathrm{g}\), which
is defined as the difference of the ionization energy and the electron
affinity. In DFT, the gap is not simply the difference between the
Kohn-Sham (KS) eigenvalues of the highest occupied molecular orbital
(HOMO) and the lowest unoccupied molecular orbital (LUMO). Instead, it
is given by~\cite{Per1982PRL,*Per1983PRL,*Sha1983PRL}
\begin{equation}
  \label{eq:gap}
  E_\mathrm{g} = \epsilon_\mathrm{LUMO}-\epsilon_\mathrm{HOMO} + \Delta_\mathrm{xc}\ ,
\end{equation}
where \(\epsilon_\mathrm{HOMO}\) and \(\epsilon_\mathrm{LUMO}\) are
the HOMO and LUMO KS eigenvalues, respectively, and
\(\Delta_\mathrm{xc}\) is the derivative discontinuity (DD) of the XC
energy with respect to the particle number N,
\begin{equation}
\label{eq:dxcdef}
\Delta_\mathrm{xc} = V_\mathrm{xc}(N^+) - V_\mathrm{xc}(N^-)\ .
\end{equation}
For the local density approximation (LDA) and many generalized
gradient approximations (GGA), the DD is zero~\cite{Toz2003MP}. In
these approximations the predicted gap is effectively the KS gap,
which severely underestimates the experimental value. Even for
functionals that are discontinuous with the particle number, the DD is
not simple to
calculate~\cite{Cha1999JCP,All2002MP,Gru2006JCP,Hel2007EPL,*Hel2009PRA,*Lat2010ZPC}. Alternatively,
XC approximations have been proposed where the KS gap is directly used
to predict the
gap~\cite{Zhe2011PRL,*Mar2011PRB,*Dea2010PRB,*Tra2009PRL,*Son2007JCP,*Hey2003JCP,*Tou2004PRA}
avoiding the calculation of the DD.

In this article, we present an XC functional for finite systems that,
with similar computational cost as a LDA or GGA calculation, has the
right asymptotic limit for low density regions and directly provides
the value of the DD. Hence, the proposed functional can predict the
fundamental gap for an atom or molecule as a ground state property.

The approach that we advocate is not based on increasing the number of
functional variables, but on changing the way that the XC potential
is described: we consider the XC potential as an electrostatic
potential, generated by a fictitious XC charge density. In contrast to
directly modelling the potential, the XC density becomes the quantity
to approximate as a functional of the electronic density \(n\).

Given a XC potential \(V_{\mathrm{xc}}\), we define the XC density
\(n_{\mathrm{xc}}\) by the Poisson equation (atomic units are used throughout)
\begin{equation}
  \label{eq:poisson_xc}
  \nabla^2V_{\mathrm{xc}}(\vec{r}) = -4\pi\,
  n_{\mathrm{xc}}(\vec{r})\ ,
\end{equation}
with the boundary condition \(V_{\mathrm{xc}}(\vec{r}\to\infty)=0\).

The justification for this approach comes from an important property
of the XC potential: the so-called asymptotic
limit~\cite{Alm1985PRB,Van1994PRA},
\begin{equation}
  \label{eq:limit}
  V_{\mathrm{xc}}(\vec{r}) \approx -\frac{1}{\left|\vec{r}\right|}\quad(\left|\vec{r}\right|\to\infty)\ ,
\end{equation}
The LDA and most GGAs do not obey the asymptotic limit condition. In
part, this common deficiency can be explained simply.
In regions that are spatially far away from the system, the density
and its derivatives decay exponentially to zero~\cite{Van1994PRA}. It
is difficult to use local values of the density and its derivatives to
reproduce a field that decays to zero much more slowly.

For the XC density the
asymptotic limit implies two simple conditions: normalization and
localization. Formally,
\begin{subequations}
\label{eq:nxc_limit}
\begin{align}
  n_{\mathrm{xc}}(\vec{r}) &= 0\quad(\left|\vec{r}\right|\to\infty)\
  ,\label{eq:nxc_limit_a}\\
  \int\mathrm{d}\vec{r}\, n_\mathrm{xc}(\vec{r}) &=
  -1\label{eq:nxc_limit_b}\ ,
\end{align}
\end{subequations}
(proof in supp. mat.~\cite{suppmat}). These constraints are similar to
the ones for the electronic density. So, in principle, it is simple to
construct a local or semi-local density functional
\(n_\mathrm{xc}[n]\) with the proper asymptotic limit. In fact, a
direct example of this XC density based approach is the functional
\(n_\mathrm{xc}[n](\vec{r}) = -n(\vec{r})/N\), which yields the
Fermi-Amaldi XC potential~\cite{Fer1934ACI,*Par1995PRA} and provides
an accurate approximation of \(V_\mathrm{xc}\) in the asymptotic
regime~\cite{Zha1994PRA,Ume2006PRA}.

\begin{figure}
  \centering
  \includegraphics[width=\columnwidth]{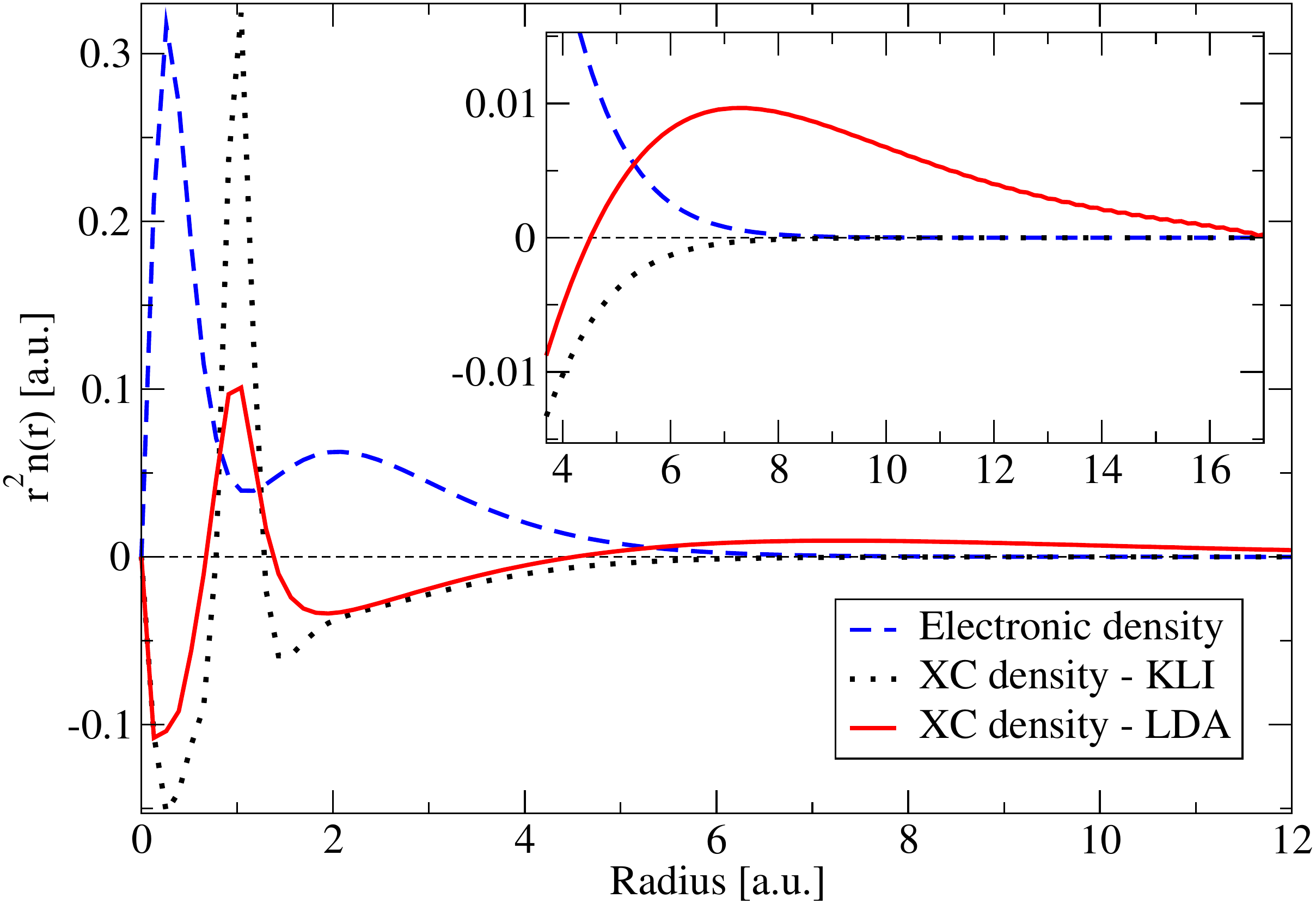}  
  \caption{KLI electronic density and KLI and LDA XC densities for the
    beryllium atom. The inset details the asymptotic region, where the
    KLI XC density shows the correct behavior. The LDA XC density is
    incorrect in this region, with a positive part that screens the
    charge in the central region.}
    \label{fig:nxc}
\end{figure}

It is illustrative to see what the XC density looks like for standard
DFT functionals. For a given XC potential, \(n_{\mathrm{xc}}\) can be
calculated analytically (see supp. mat.~\cite{suppmat}), but in
practice it is simpler to evaluate Eq.~(\ref{eq:poisson_xc})
numerically. In Fig.~\ref{fig:nxc}, we compare \(n_\mathrm{xc}\) for
LDA and exact exchange in the Krieger-Li-Iafrate
(KLI)~\cite{Kri1992PRA} approximation. As can be seen from the figure,
for large radius the KLI XC density correctly goes to zero faster than
the electronic density. On the other hand, the LDA XC density becomes
positive. This positive XC charge screens the XC charge in the central
region making the total XC charge zero and therefore causing the
potential to decay exponentially.

{ As a first application of our approach, we propose a
  correction method to enforce the proper asymptotic limit for
  functionals that do not have it by construction. Given a certain
  potential \(\bar{V}_\mathrm{xc}\), we calculate the associated
  \(\bar{n}_\mathrm{xc}\) from Eq.~(\ref{eq:poisson_xc}). To this XC
  density we apply the correction procedure, that generates a
  corrected XC density \(n^c_\mathrm{xc}\). In turn, the corrected XC
  density is used to reconstruct the corrected XC potential
  \(V^c_\mathrm{xc}\) by solving Eq.~(\ref{eq:poisson_xc}).  }

The correction procedure for \(\bar{n}_\mathrm{xc}\) enforces it to be
localized by setting it to zero when the local value of electronic
density is below a certain threshold \(\eta\). This simple procedure
can be written as a correction term \(\Delta n_\mathrm{xc}\) to be
added to the XC density of the original functional,
\begin{equation}
  \label{eq:functional}
  \Delta n_\mathrm{xc}[n](\vec{r}, \eta) = \
  \left\{
    \begin{array}{ll}
      0 & 
      \mbox{if } n(\vec{r}) \geq \eta\\
      -\bar{n}_\mathrm{xc}[n](\vec{r}) & \mbox{if } n(\vec{r}) < \eta
    \end{array}
  \right.\ .
\end{equation}
To determine the parameter \(\eta\), for each density we obtain an
optimized value \(\eta_0\) that tries to enforce Eq.~(\ref{eq:nxc_limit_b}).
First, we define the total XC charge as a function of \(\eta\)
\begin{equation}
  \label{eq:qxc}
  q_\mathrm{xc}(\eta) =
  \int\mathrm{d}\vec{r}\,\left\{\bar{n}_\mathrm{xc}[n](\vec{r}) + \Delta
    n_\mathrm{xc}[n](\vec{r}, \eta)\right\}\ .
\end{equation}
Ideally, from Eq.~(\ref{eq:nxc_limit_b}), we need to find \(\eta_0\)
such that \(q_\mathrm{xc}(\eta_0)=-1\). However, there is no guarantee
about the existence or uniqueness of \(\eta_0\). So we choose
\(\eta_0\) such that \(q_\mathrm{xc}(\eta_0)\) has the closest value
to -1, with \(\eta_0\) restricted to be smaller than the first minimum
of \(q_\mathrm{xc}(\eta)\). (See supp. mat.~\cite{suppmat}.)

When \(q_\mathrm{xc}(\eta_0)\neq-1\), Eq.~(\ref{eq:nxc_limit_b}) is
still not satisfied, so we rescale the correction by
\(\left|q_\mathrm{xc}(\eta_0)\right|^{-1}\). The final expression for the
XC density of the corrected functional is
\begin{equation}
  \label{eq:scaldiff}
  n^c_{\mathrm{xc}}(\vec{r}) = \bar{n}_{\mathrm{xc}}(\vec{r}) +
  \frac1{\left|q_\mathrm{xc}(\eta_0)\right|}\Delta
  n_{\mathrm{xc}}(\vec{r}, \eta_0)\ .
\end{equation}
This rescaling form guarantees that Eq.~(\ref{eq:nxc_limit_b}) is
satisfied, and that the original XC potential is changed as little as
possible in the central region (the region where \(n\ge\eta_0\)).

In theory, it is only the exchange term that is responsible for the
long range behavior, due to the much faster decay of the correlation
term~\cite{Van1994PRA}, therefore, the correction can be applied
either to the exchange potential or the full XC potential. In this
work, we apply it to the exchange part of the LDA functional, and we
call the combination of the corrected LDA exchange and LDA correlation
(in the Perdew-Wang form~\cite{Per1992PRBb}) the corrected exchange
density LDA (CXD-LDA). For spin-polarized systems, the correction is
calculated for the spin-unpolarized potential using the total
density. Then the difference between the corrected potential and the
original one is added to the XC potential for each spin component.

To test the CXD-LDA functional, we performed calculations for atoms,
hydrogen to strontium, and for a set of small molecules. We
implemented the correction procedure in the APE~\cite{Oli2008CPC} and
Octopus~\cite{Cas2006PSSB} codes. We find that the optimized value of
\(\eta_o\) changes significantly for different systems. For most atoms
\(q_\mathrm{xc}(\eta_0)\neq-1\) while for all the tested molecules
\(q_\mathrm{xc}(\eta_0)=-1\) (see supp. mat.~\cite{suppmat}). The
numerical cost of a self-consistent solution using the correction is
similar to the LDA calculation (see supp. mat. for
details~\cite{suppmat}).

\begin{figure}
  \centering
  \includegraphics[width=0.95\columnwidth]{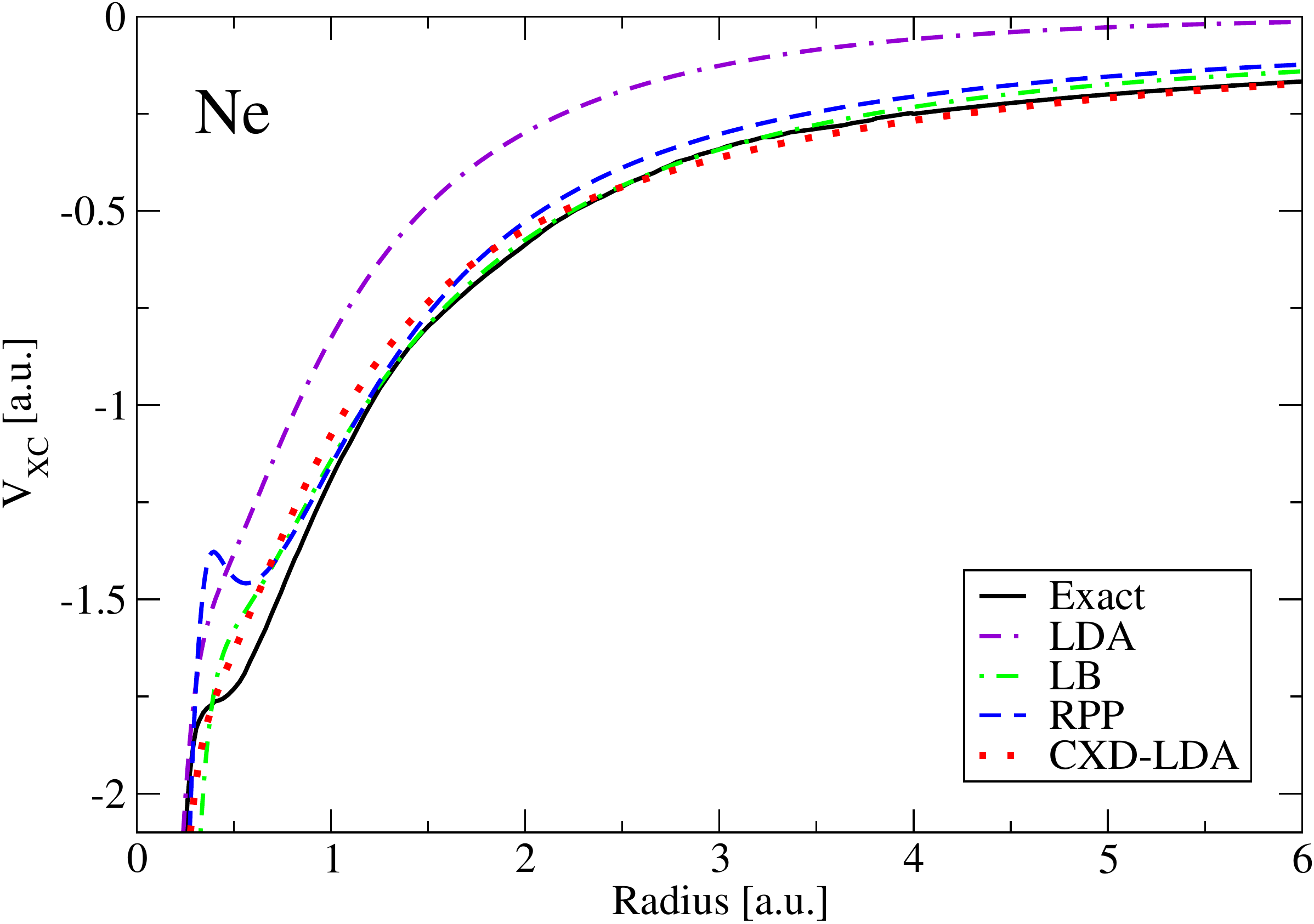}  
  \caption{CXD-LDA potential for neon. Comparison with the LDA,
    LB~\cite{Van1994PRA}, RPP~\cite{Ras2010JCP} and the
    exact~\cite{Zha1994PRA} potentials.}
  \label{fig:vxc}
\end{figure}

In Fig.~\ref{fig:vxc}, we show the CXD-LDA potential for Ne compared
with an accurate approximation to the exact
potential~\cite{Zha1994PRA}, the original LDA functional and two
functionals that have the correct asymptotic behavior: the van
Leeuwen-Baerends (LB) GGA~\cite{Van1994PRA} and the
R\"{a}s\"{a}nen-Pittalis-Proetto (RPP) meta-GGA~\cite{Ras2010JCP}.

For atoms it is simple to understand the effect of the correction
procedure. Due to the spherical symmetry and the
monotonically-decreasing density, the correction XC charge \(\Delta
n_\mathrm{xc}\) is a spherical shell. By Newton's shell theorem, the
correction potential will be constant in the central region. Outside,
the correction will decay close to \(-1/r\).  We can expect this
behavior to be similar for more complex systems if the
\(n(\vec{r})=\eta_0\) surface is close to a sphere (see supp. mat.~\cite{suppmat}).

{ The shape of the correction is similar to the one
  proposed by Casida and Salahub~\cite{Cas1998IJQC,Cas2000JCP}, who
  argument that a shift of the XC potential in the central region is
  necessary to fix the asymptotic limit of the LDA
  potential. Moreover, they show that the shift is related to the DD
  of the energy with respect to the particle number. As in our method
  the shift appears naturally from imposing the asymptotic limit, we
  can obtain the value of the DD.

  To obtain the relation between the DD and the shift, we assume that
  a potential \(\bar{V}_\mathrm{xc}\), which lacks the DD,
  approximates the XC potential averaged over the
  discontinuity~\cite{All2002MP,Toz2003MP}. This is
  \begin{equation}
    \label{eq:vxcav}
    \bar{V}_\mathrm{xc} = \frac12\left[V_\mathrm{xc}(N^+) +
      V_\mathrm{xc}(N^-)\right] \ .
  \end{equation}
  Using Eq.~(\ref{eq:dxcdef}), immediately follows that
  \begin{equation}
    \label{eq:vxcdxc}
    V_\mathrm{xc}(N^-) = \bar{V}_\mathrm{xc} - \frac12\Delta_\mathrm{xc}\ .
  \end{equation}
  
  By imposing the asymptotic limit of Eq.~\ref{eq:limit}, our
  corrected potential is approximating
  \(V_\mathrm{xc}(N^-)\)~\footnote{For a continuous number of
    particles, the limit of Eq.~(\ref{eq:limit}) is generalized to
    \(-f/r\) where \(f\) is the occupation of the
    HOMO~\cite{Cas2000JCP}.}. Therefore, we can obtain the value of
  the DD from Eq.~(\ref{eq:vxcdxc}). For practical calculations we
  average the change of the XC potential due to the correction over
  the central region
  \begin{equation}
    \label{eq:dxc}
    \Delta_\mathrm{xc}=-\frac2\Omega\int_{n(\vec{r})\ge\eta}\mathrm{d}\vec{r}\left[V^c_\mathrm{xc}(\vec{r})-\bar{V}_\mathrm{xc}(\vec{r})\right]\ ,
  \end{equation}
  where \(\Omega\) is the volume of the central region. This
  expression for the DD can be calculated directly from the correction
  process as a ground-state property. An alternative, but less
  practical, method for the calculation of the DD is detailed in
  supp. mat.~\cite{suppmat}.

}

\begin{table}
  \centering
  \caption{Comparison of the calculated derivative
    discontinuity with theoretical and experimental results. \(^a\)
    Experimental gap values from
    Ref.~\cite{NIST}. \(^b\) Ensemble spin DFT results by
    Chan~\cite{Cha1999JCP}. Values in atomic units.}
  \label{tab:deltaxc}
  \begin{tabular*}{0.8\columnwidth}{@{\extracolsep{\fill}} ccccc}
    \hline\hline
    Atom & Experimental\(^a\) & ESDFT\(^b\) & CXD-LDA \\
     \hline
     B &  0.295 & 0.270 & 0.284 \\
     C &  0.367 & 0.342 & 0.337 \\
     O &  0.447 & 0.404 & 0.435 \\
     F &  0.515 & 0.478 & 0.467 \\
    \hline\hline
   \end{tabular*}
\end{table}

In Table~\ref{tab:deltaxc}, we compare the DD obtained with
Eq.~(\ref{eq:dxc}) with the values reported by Chan~\cite{Cha1999JCP}
from ensemble DFT (with XC potentials obtained from wave-function
methods). We also compare it with the experimental value of the gap,
that for these open-shell atoms is equal to the DD since the KS gap is
zero. The three sets present a remarkable agreement, with our results
being smaller that the experimental values by less than 10\%.

\begin{figure}
  \centering
  \includegraphics[width=\columnwidth]{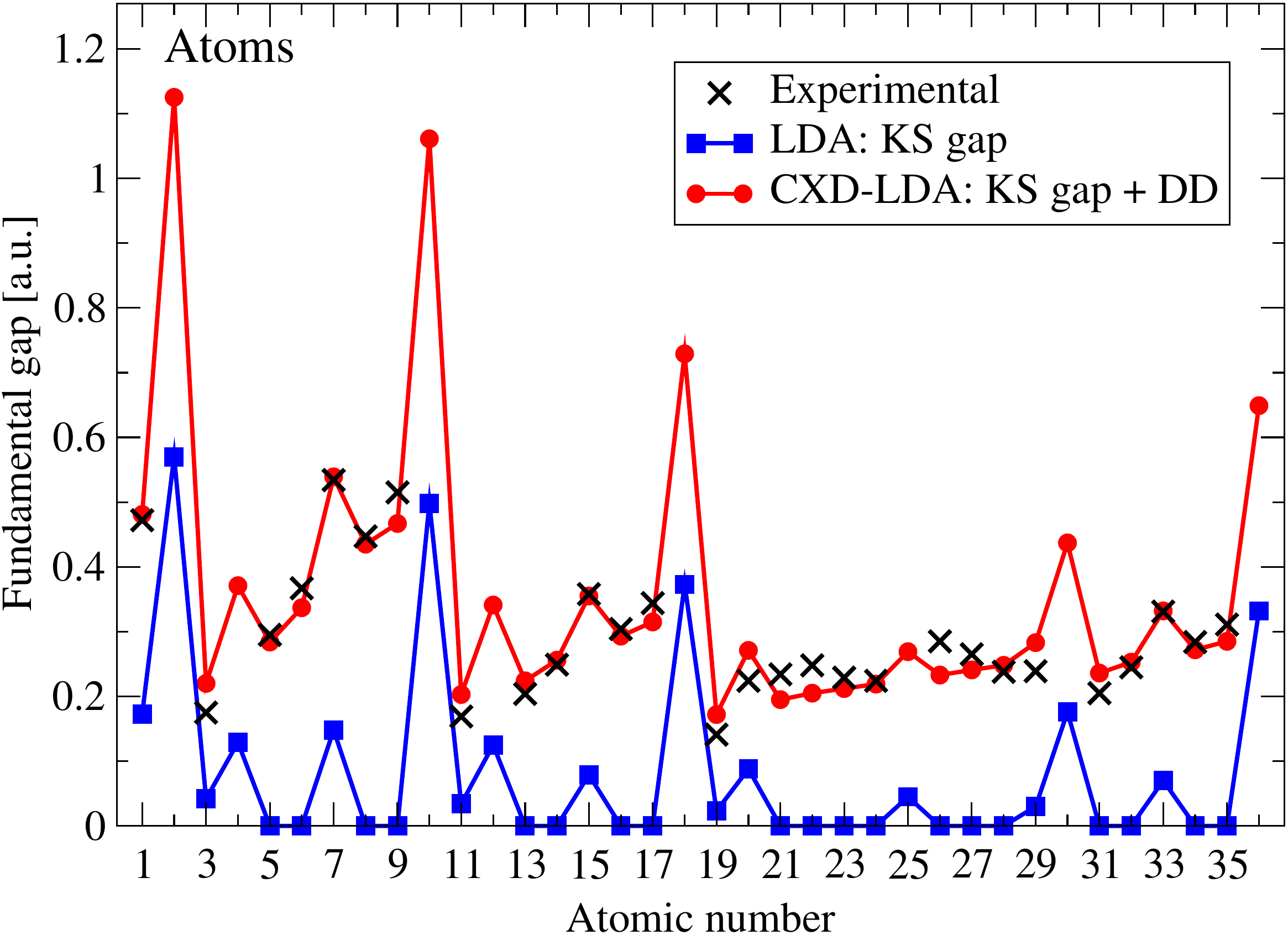}  
  \includegraphics[width=\columnwidth]{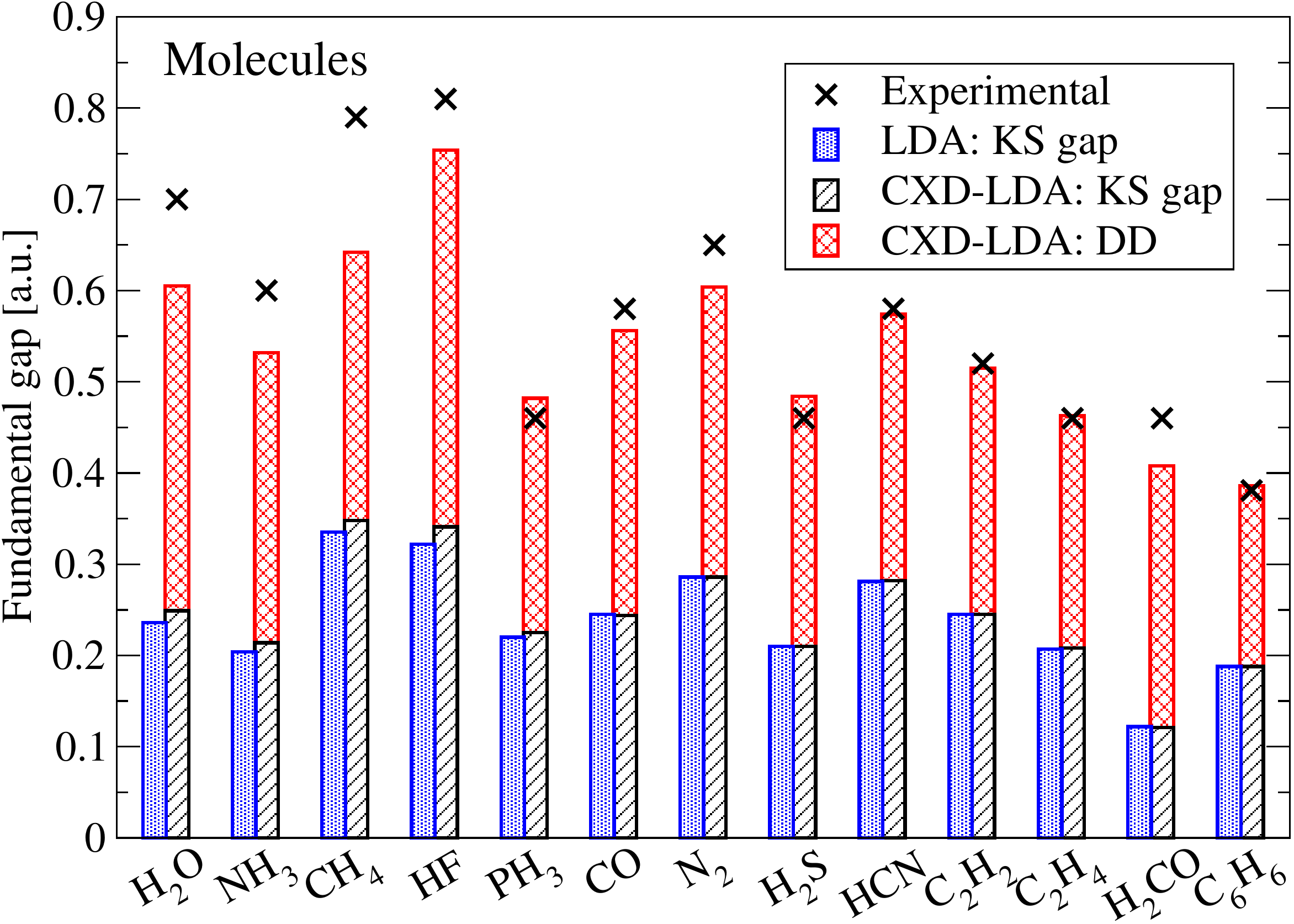}  
  \caption{Comparison of the CXD-LDA gap with LDA and available
    experimental results. The CXD-LDA gap has contributions from the
    Kohn-Sham (KS) gap and the derivative discontinuity (DD). Result
    for atoms (top) from H to Kr, experimental values from
    Ref.~\cite{NIST}, and a set of molecules (bottom), experimental
    values compiled in Ref.~\cite{All2002MP}, except for
    C\(_6\)H\(_6\)~\cite{NIST,Bur1987JCP}. (Data in
    supp. mat.~\cite{suppmat}.)}
  \label{fig:gap}
\end{figure}

To investigate further the quality of corrected functional and its DD,
we compare the calculated gap with the LDA KS gap and the experimental
gap, for our set of atoms and molecules. The results are plotted in
Fig.~\ref{fig:gap}. The KS gap of the corrected functional is close to
the LDA one~\footnote{Better LDA values could be obtained from
  differences in total energies~\cite{Oli2010JCTC}.} and far from the
experimental value. Once we add the DD, however, the results are
closer to the experiment, with an average error of \(11\%\) for atoms
and \(7\%\) for molecules.

\begin{figure}
  \centering
  \includegraphics[width=\columnwidth]{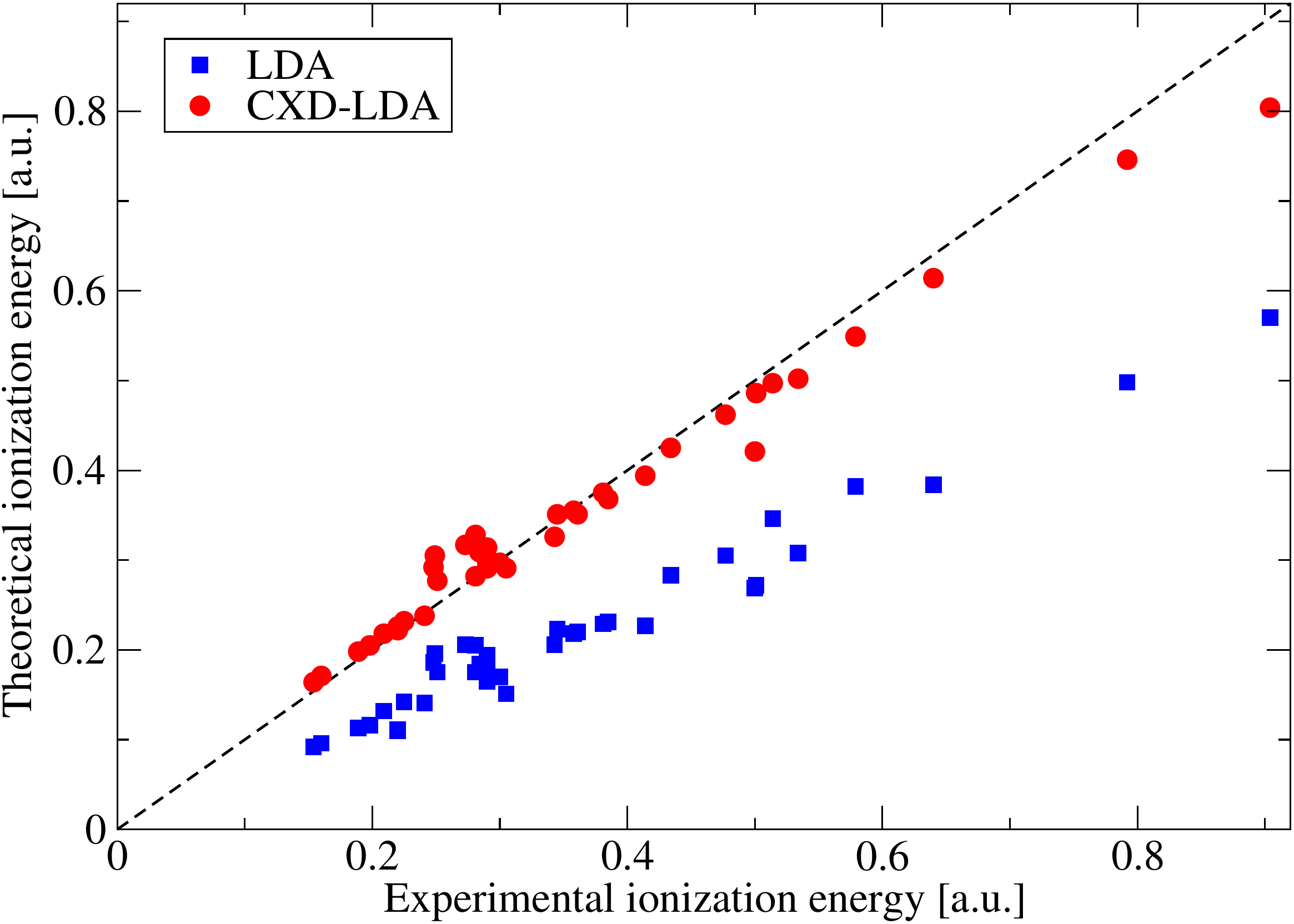}
  \caption{LDA and CXD-LDA ionization energy as a function of the
    experimental value. Atoms from H to Sr. (Data in
    supp. mat.~\cite{suppmat}.)}
  \label{fig:atomion}
\end{figure}

While the correction has little effect on the KS gap, it changes the
KS eigenvalues. This can be seen in the ionization energy (I), which in
DFT is given by \(-\epsilon_\mathrm{HOMO}\)~\cite{Alm1985PRB}. In
Fig.~\ref{fig:atomion}, we plot I for the LDA and CXD-LDA functionals
as a function of the experimental value. In Table~\ref{tab:atomion},
we compare the deviation from experimental results for atoms with
other XC functionals that have the proper asymptotic limit: LB, RPP,
and KLI with Colle-Salvetti correlation~\cite{Col1975TCA}
(KLI-CS). The correction procedure improves considerably the LDA
results, with similar accuracy to other long range XC potentials.

\begin{table}
  \centering
  \caption{Mean absolute error in the ionization energy for atoms
    (He to Ar). Comparison of CXD-LDA
    with LDA, LB~\cite{Van1994PRA}, RPP~\cite{Ras2010JCP} and
    KLI-CS~\cite{Kri1992PRA,Col1975TCA} functionals.
    \(^a\)Results from Ref.~\cite{Oli2010JCTC}. \(^b\)Results from
    Ref.~\cite{Gra1995CPL} given in Ref.~\cite{Oli2010JCTC}. (Data in supp. mat.~\cite{suppmat}.)}
  \label{tab:atomion}
  \begin{tabular*}{0.8\columnwidth}{@{\extracolsep{\fill}} cccccc}
    \hline\hline
    LDA &  LB\(^a\) &  RPP\(^a\) & KLI-CS\(^b\) &
    CXD-LDA \\
    \hline
    41\% & 3.7\% & 7.4\% & 5.7\% & 4.1\%\\
    \hline\hline
  \end{tabular*}
\end{table}

In summary, we have introduced a new auxiliary quantity, the XC
density, to construct approximations for the XC potential. Based on an
exact condition that the XC potential must fulfill and basic notions
of electrostatics, we have presented a correction method for any
previously proposed XC potential.

Additionally, the correction procedure allows for the direct
calculation of the DD of the XC energy, which can be used to directly
predict the fundamental gap as a ground-state property. Moreover, our
approach allows for a routine computation of the DD as a practical
method for the prediction of the gap.

The proposed potential is a pure functional of the electronic density
with a certain degree of non-locality included by the optimization of
\(\eta_0\) and by the Poisson equation. The correction procedure does
not depend on any empirical or globally adjusted parameter.

Since the basis for our method is the correction of the XC potential
in the asymptotic region, it is not directly applicable to crystalline
systems. However, the concepts of the XC density and the DD as a
potential shift are still valid. Therefore, it might be possible to
generalize the method to solids, where the determination of accurate
gaps is one of the main challenges for DFT.

\begin{acknowledgments}
We acknowledge support from the US National Science Foundation (DMR-0934480).
\end{acknowledgments}

\bibliography{biblio}

\clearpage

\appendix

\section{Supplementary Material}

\subsection{Analytic expression for the XC density}

Given an XC potential, the XC density is given by 
\begin{equation}
  \label{smeq:poisson_xc}
  n_{\mathrm{xc}}[n](\vec{r}) =
  -\frac1{4\pi}\nabla^2V_{\mathrm{xc}}[n](\vec{r})\ ,
\end{equation}
by using the chain rule for functional derivatives
\begin{multline}
  \label{smeq:nxc_analytic}
  n_{\mathrm{xc}}[n](\vec{r}) =
  -\frac1{4\pi}\left[\int\mathrm{d}\vec{r}'f_{\mathrm{xc}}[n](\vec{r},\vec{r}')\nabla^2n(\vec{r}')\right.+\\
  \left.+\int\mathrm{d}\vec{r}'\mathrm{d}\vec{r}''k_{\mathrm{xc}}[n](\vec{r},\vec{r}',\vec{r}'')\,\vec{\nabla}n(\vec{r}')\cdot\vec{\nabla}n(\vec{r}'')\right]\
  ,
\end{multline}
where \(f_{\mathrm{xc}}\) and \(k_{\mathrm{xc}}\) are, respectively,
the first and second functional derivatives of \(V_{\mathrm{xc}}\)
with respect to the density. Interestingly,
Eqs.~(\ref{smeq:poisson_xc})~and~(\ref{smeq:nxc_analytic}) give a recipe
to reconstruct an XC potential from its first two functional
derivatives.

\subsection{Proof of the conditions for the XC density}

The asymptotic condition for \(V_\mathrm{xc}\) implies that there
exists a certain \(r_c\) such that
\begin{equation}
  \label{smeq:limit}
  V_{\mathrm{xc}}(\vec{r}) = -\frac{1}{\left|\vec{r}\right|}\quad(\left|\vec{r}\right|\geq\vec{r_c})\ .
\end{equation}
If we apply the Laplacian over this equality, we get
\begin{equation}
  \label{smeq:limita}
  n_{\mathrm{xc}}(\vec{r}) = 0\quad(\left|\vec{r}\right|\geq\vec{r_c})\ , 
\end{equation}
the localization condition.

For the normalization condition, we integrate
Eq.~(\ref{smeq:poisson_xc}) over a spherical volume of radius \(r_c\)
and we use Gauss's theorem over the right-hand side
\begin{equation}
  \label{smeq:int}
  \int_{r_c\geq r}\mathrm{d}\vec{r}\,n_{\mathrm{xc}}(\vec{r}) =
  -\frac1{4\pi}\oint_{r_c=r}\mathrm{d}\vec{A}\cdot\nabla V_{\mathrm{xc}}(\vec{r}).
\end{equation}
From Eq.~(\ref{smeq:limit}), \(\nabla V_{\mathrm{xc}}[n]=\hat{r}/r^2\)
which is constant over the sphere, so the integral on the right is
just \(1/r_c^2\) times the surface of the sphere, \(4\pi r_c^2\). The
integral on the left hand size can be extended to whole space due to
Eq.~(\ref{smeq:limita}), so finally
\begin{equation}
  \label{smeq:nxc_limit_b}
  \int\mathrm{d}\vec{r}\,n_{\mathrm{xc}}(\vec{r}) = -1\ .
\end{equation}
\subsection{Optimization of the \(\eta\) parameter}

\begin{figure}
  \centering
  \includegraphics[width=1.0\columnwidth]{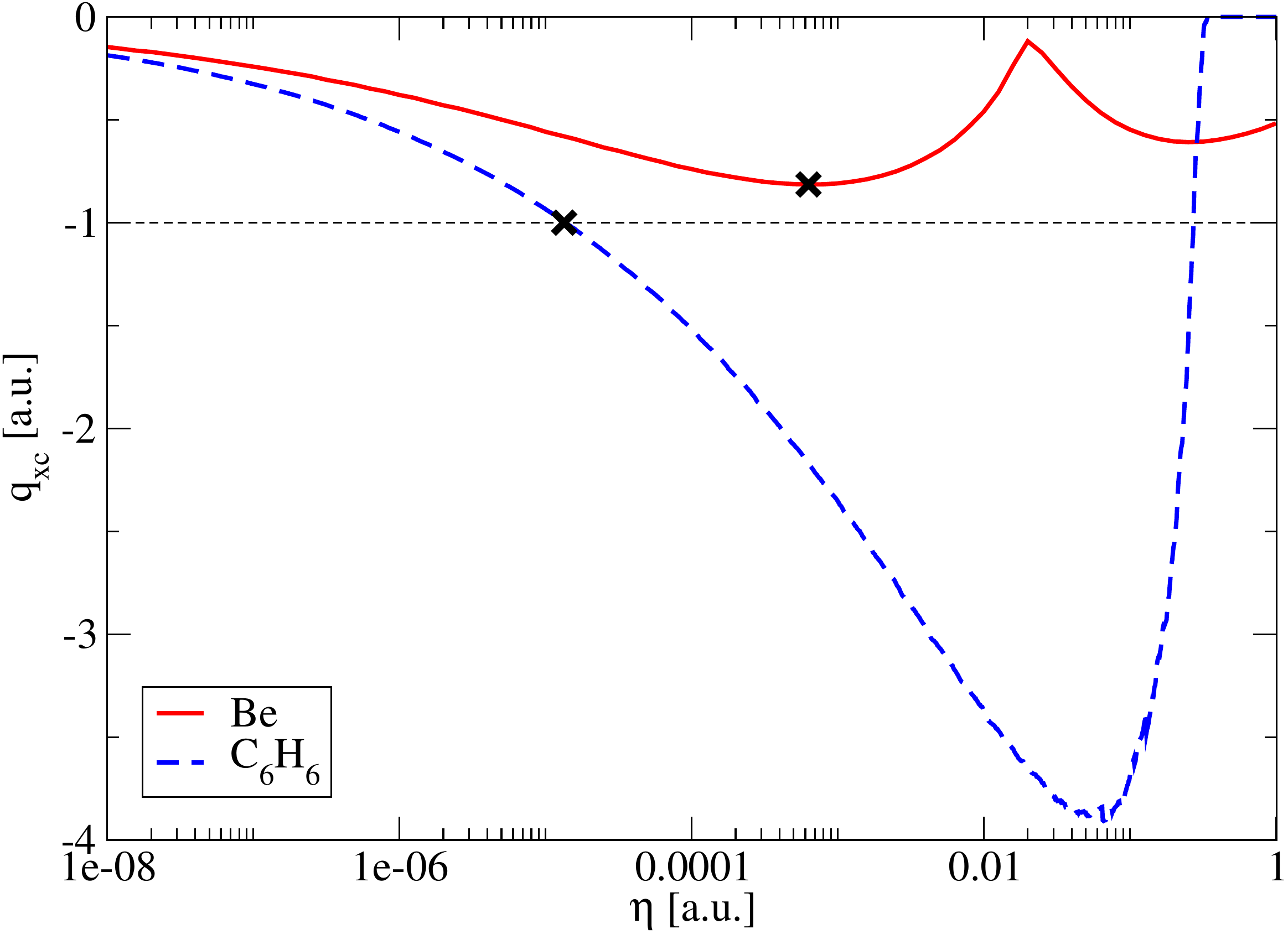}  
  \caption{The \(q_\mathrm{xc}(\eta)\) function, Eq.~(\ref{smeq:qxc}),
    for LDA on atomic Beryllium atom and C\(_6\)H\(_6\). The crosses
    mark the position of \(\eta_0\), the optimal value of \(\eta\).}
  \label{smfig:qxc}
\end{figure}

Given a XC density \(\bar{n}_\mathrm{xc}\), the corrected XC density
is defined as
\begin{equation}
  \label{smeq:scaldiff}
  n_{\mathrm{xc}}(\vec{r}) = \bar{n}_{\mathrm{xc}}(\vec{r}) +
  \frac1{\left|q_\mathrm{xc}(\eta)\right|}\Delta
  n_{\mathrm{xc}}(\vec{r}, \eta)\ ,
\end{equation}
with the correction XC density
\begin{equation}
  \label{smeq:functional}
  \Delta n_\mathrm{xc}[n](\vec{r}, \eta) = \
  \left\{
    \begin{array}{ll}
      0 & 
      \mbox{if } n(\vec{r}) \geq \eta\\
      -\bar{n}_\mathrm{xc}[n](\vec{r}) & \mbox{if } n(\vec{r}) < \eta
    \end{array}
  \right.\ ,
\end{equation}
and the total XC charge
\begin{equation}
  \label{smeq:qxc}
  q_\mathrm{xc}(\eta) =
  \int\mathrm{d}\vec{r}\,\left\{\bar{n}_\mathrm{xc}[n](\vec{r}) + \Delta
    n_\mathrm{xc}[n](\vec{r}, \eta)\right\}\ .
\end{equation}
The value of \(\eta\) is optimized for each density, the optimum value
\(\eta_0\) is the value that has the total charge closest to -1. Formally,
\begin{equation}
  \label{smeq:minprincpl}
  \eta_0[n]=\min_{\eta\le\eta_m}\left|q_\mathrm{xc}[n](\eta)+1\right|
\end{equation}
where \(\eta_m\) is the smallest value where \(q_\mathrm{xc}(\eta)\) has
a minimum. These conditions are necessary, since it is not always
possible to find a unique value where \(q_\mathrm{xc}(\eta)=-1\). This
can be seen in Fig.~\ref{smfig:qxc} were we show \(q_\mathrm{xc}(\eta)\)
for two systems. While for beryllium \(q_\mathrm{xc}\) never reaches
-1, for benzene there are two points where it does. The marks in the
figure indicate the value of \(\eta_0\) and \(q_\mathrm{xc}(\eta_0)\)
selected by our method.

\subsection{Corrected functional test}

To test the corrected LDA functional, CXD-LDA, we did calculations for
atomic systems from hydrogen to strontium and for a set of molecules.

\begin{figure}
  \centering
  \includegraphics[width=1.0\columnwidth]{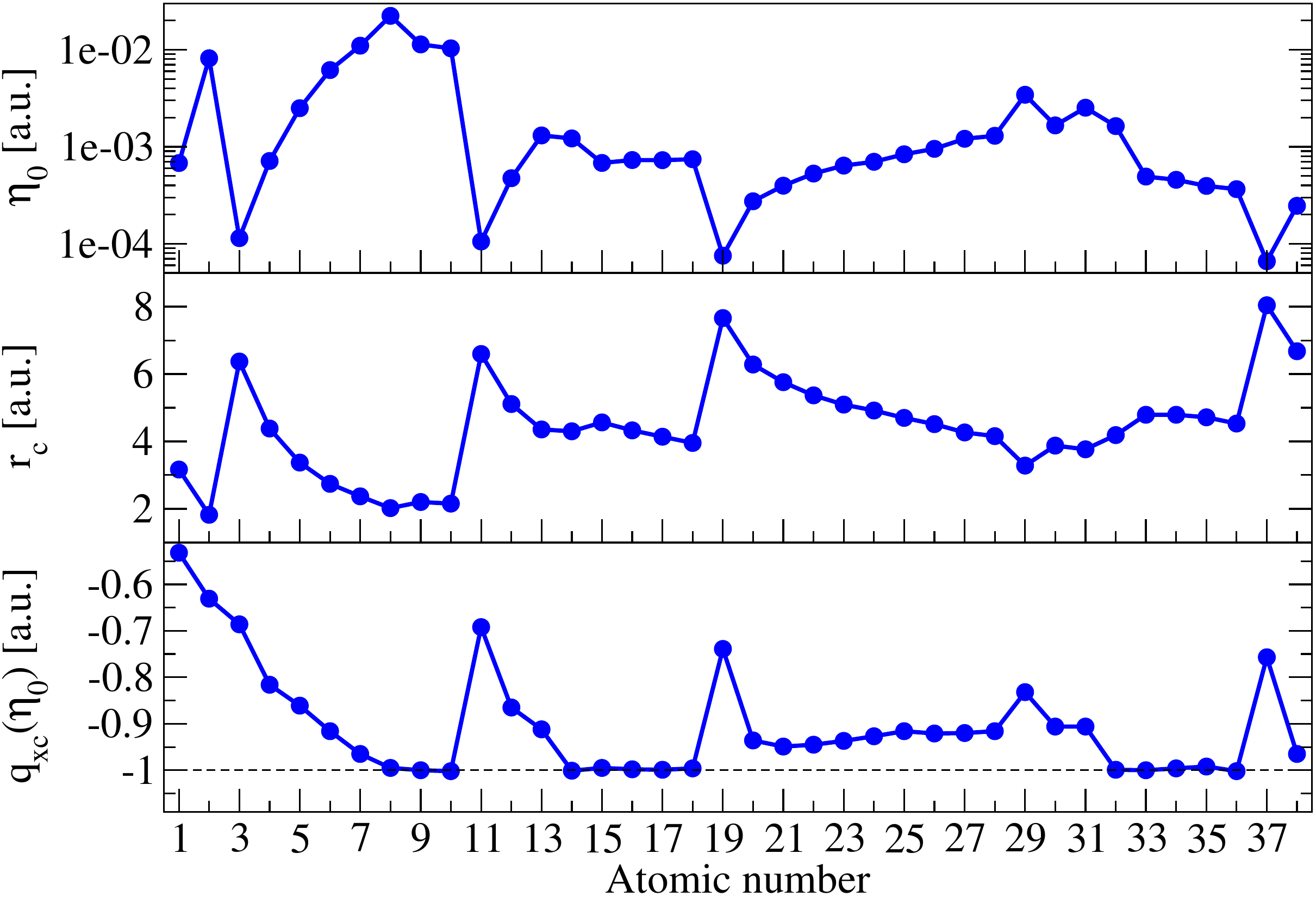}
  \caption{CXD-LDA optimized parameters for atoms. Top: the density
    threshold \(\eta_0\). Center: the radius \(r_c\) where
    \(n(r_c)=\eta_0\). Bottom: The value of the optimized total XC
    charge, \(q_\mathrm{xc}(\eta_0)\).}
  \label{smfig:param}
\end{figure}

\begin{figure}
  \centering
  \includegraphics[width=1.0\columnwidth]{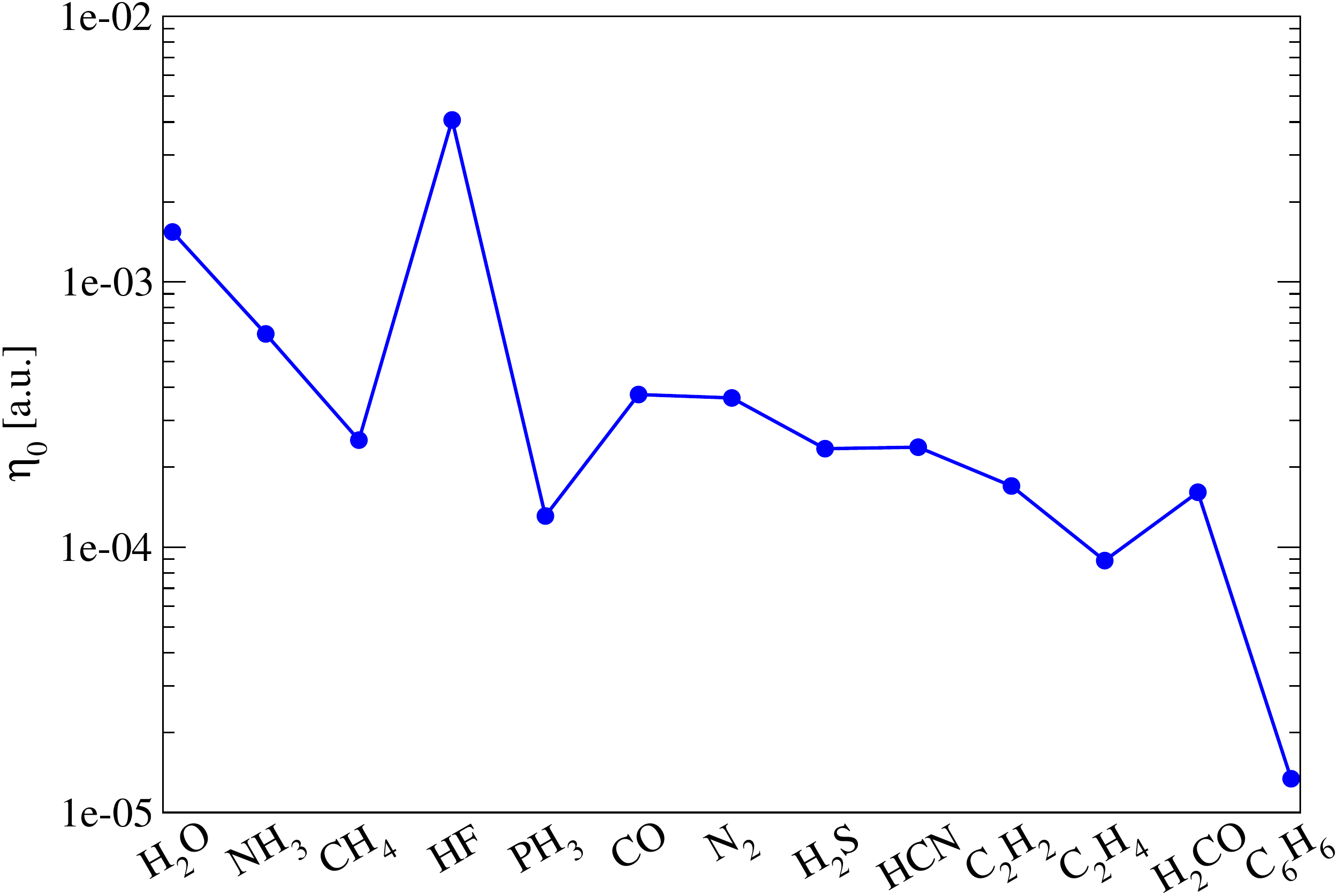}

  \caption{Optimized density threshold, \(\eta_o\), for a set of small
    molecules. The value of the total XC charge,
    \(q_\mathrm{xc}(\eta_0)\), is -1 for all cases. Values are given
    in Table~\ref{smtab:molecules}.}
  \label{smfig:molparam}
\end{figure}

\begin{figure}
  \centering
  \includegraphics[width=1.0\columnwidth]{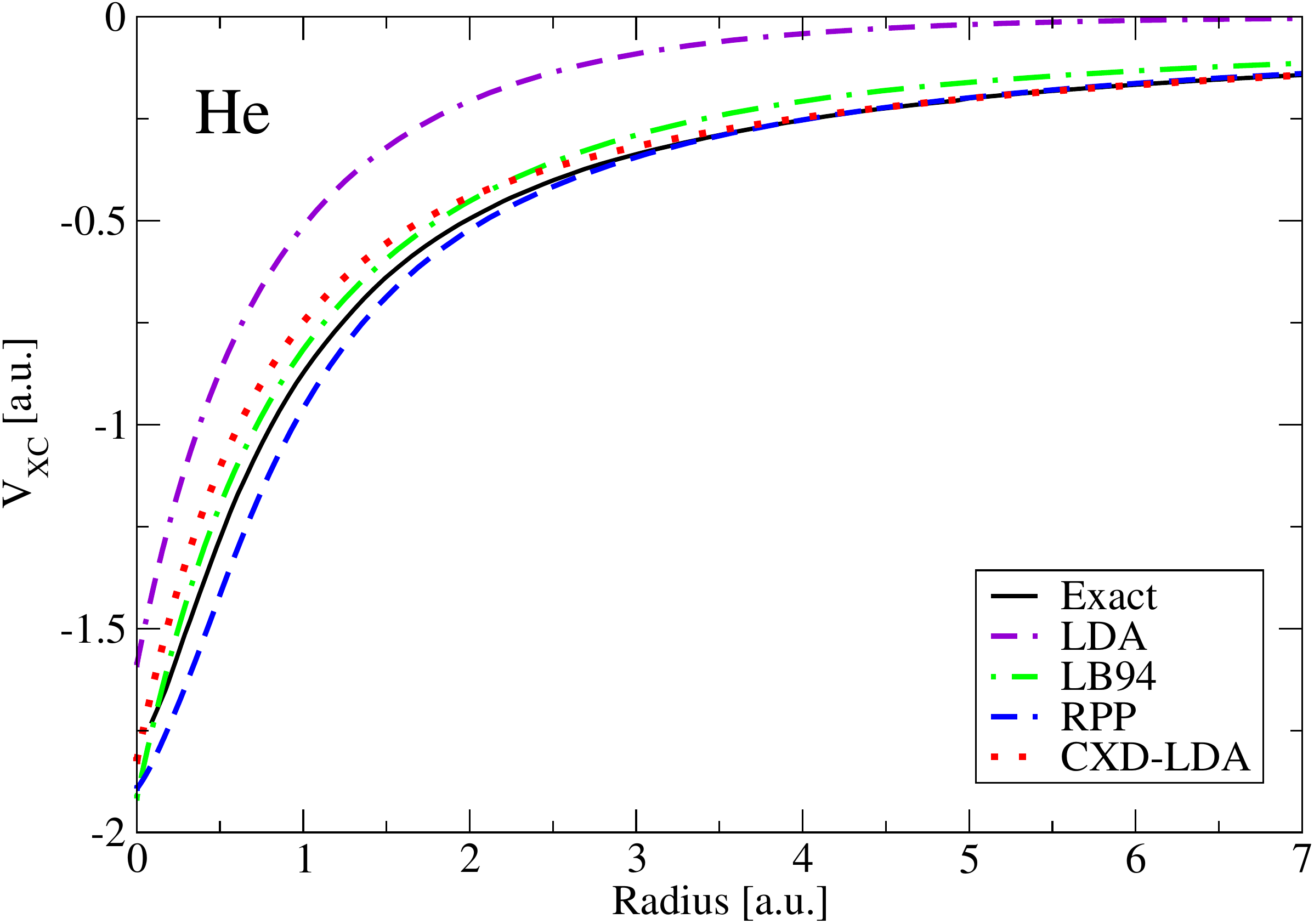}  
  \caption{CXD-LDA potential for helium. Comparison with the LDA,
    LB~\cite{Van1994PRA}, RPP~\cite{Ras2010JCP} and the exact~\cite{Zha1994PRA} potentials.}
  \label{smfig:vxc}
\end{figure}

\begin{figure*}
  \centering
  \includegraphics[width=0.95\textwidth]{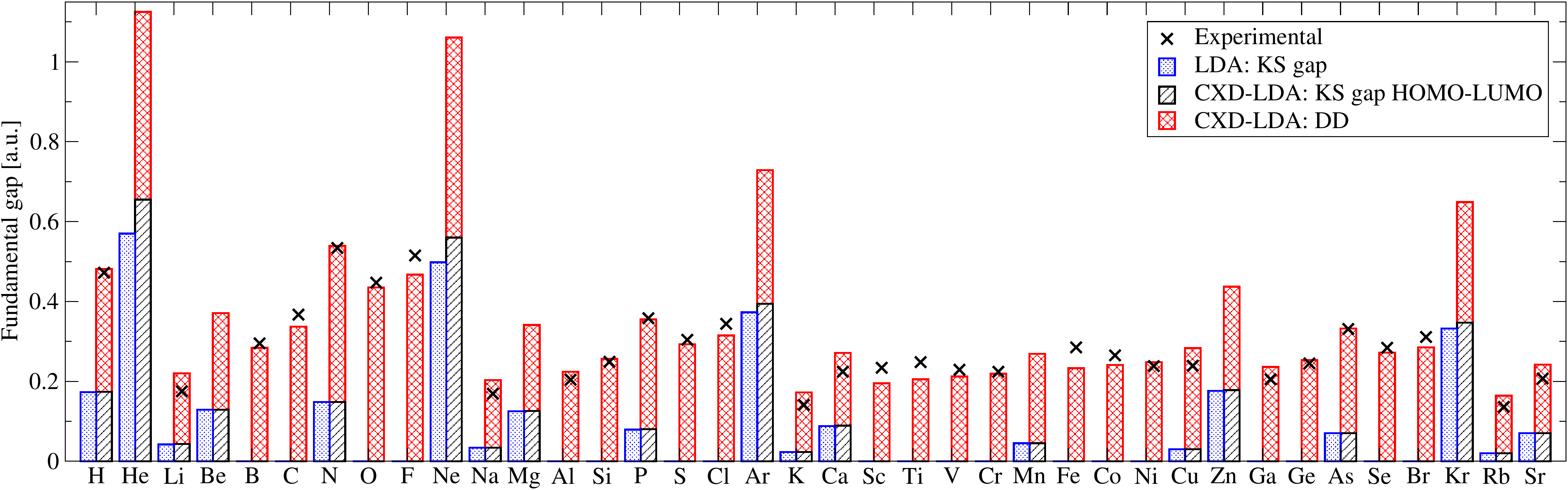}  
  \caption{Comparison of fundamental gap calculations with available
    LDA and CXD-LDA functionals with experimental results for atoms
    between H and Sr. Experimental values obtained from
    Ref.~\cite{NIST}. Values are given in Table~\ref{smtab:atoms}.}
  \label{smfig:gap}
\end{figure*}

The optimized parameter \(\eta_0\) obtained for each system is plotted
Figs.~\ref{smfig:param} and Figs.~\ref{smfig:molparam}. For atoms we also
plot the corresponding value of \(q_\mathrm{xc}(\eta_0)\) and \(r_c\),
the value where \(n(r_c)=\eta_o\), that marks the position where the
XC density is set to zero. For all tested molecules
\(q_\mathrm{xc}(\eta_0)=-1\). In Fig.~\ref{smfig:vxc}, we plot the XC
potential for He compared with other functionals. In
Fig.~\ref{smfig:gap}, we present a more detailed version of the plot for
the band gap of the atoms. Finally, we give the values for our
optimized parameters, ionization energies and gaps in
Tables~\ref{smtab:molecules} (molecules) and~\ref{smtab:atoms} (atoms).

\subsection{Shape of the correction}

\begin{figure}
   \centering
   \includegraphics[width=1.0\columnwidth]{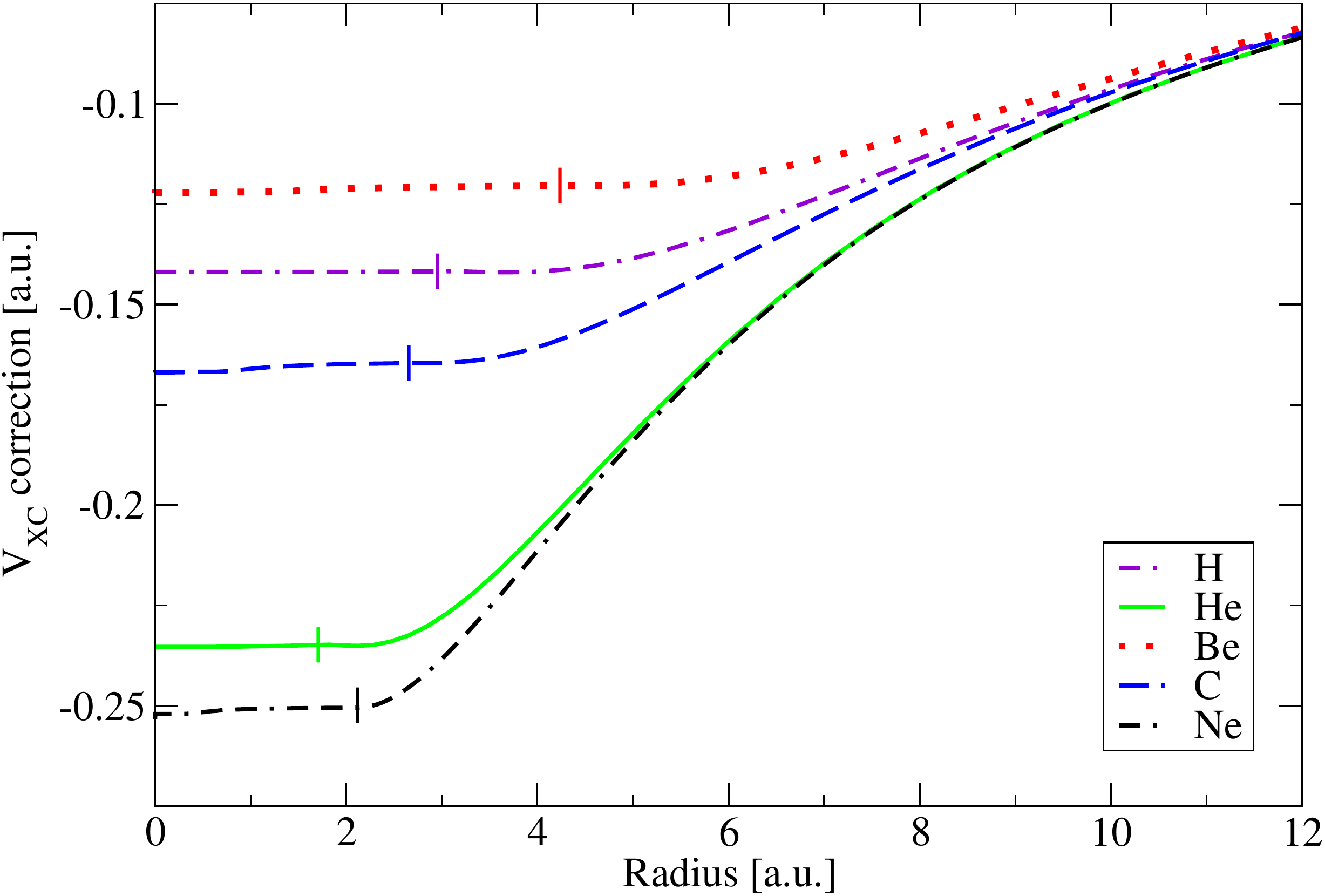}
   \includegraphics[width=1.0\columnwidth]{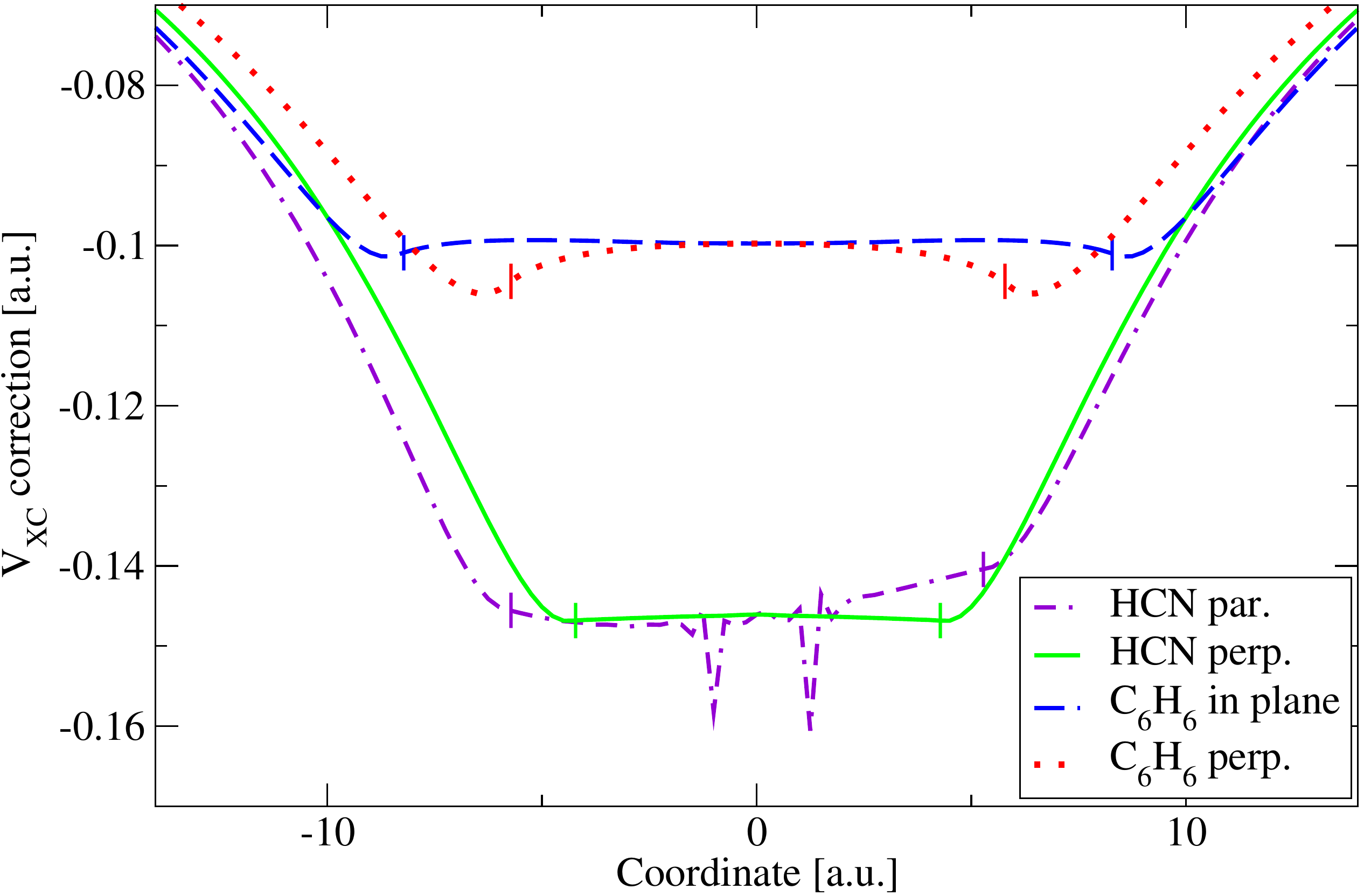}
   \caption{Correction of the LDA X potential for different atoms
     (top) and molecules (bottom). The vertical lines mark the
     position where \(n=\eta_0\). For the HCN molecule the value of
     the shift is plotted for the axis that passes by the atoms (HCN
     par.) and other axis that is perpendicular to that one (HCN
     perp.). For C\(_6\)H\(_6\) one axis in the plane of the molecule
     (C\(_6\)H\(_6\) in plane) and the other is perpendicular to that
     plane (C\(_6\)H\(_6\) perp.).}
   \label{smfig:correction}
 \end{figure}

In Fig.~\ref{smfig:correction} we plot the difference of the corrected
XC potential and the original one for 5 atoms and 2 molecules. For
atoms the correction has the shape expected from electrostatic
principles. For molecules, the correction in the central region is not
exactly constant. This is expected as the \(n=\eta_0\) surface is not
necessarily a perfect sphere. However the deviation from a constant
shift is not large and we can still obtain a representative value from
the average. Near the atomic positions there are some peaks due to
numerical error in the finite difference calculation of the Laplacian,
but these are a few points that do not influence considerably the
average.

\subsection{Alternative calculation of the derivative discontinuity}

We propose to calculate the derivative discontinuity (DD) from the
shift of the potential due to the correction procedure
\begin{equation}
  \label{smeq:dxc}
  \Delta_\mathrm{xc}^V=-\frac2\Omega\int_{n(\vec{r})\ge\eta}\mathrm{d}\vec{r}\left[V_\mathrm{xc}(\vec{r})-\bar{V}_\mathrm{xc}(\vec{r})\right]\ .
\end{equation}
Alternatively, the DD can be obtained by comparing the results of the
corrected XC potential with the results of the uncorrected
functional. In principle, in DFT the ionization energy is given by the
highest occupied Kohn-Sham eigenvalue
\begin{equation}
  \label{smeq:ehomo}
  \epsilon_\mathrm{HOMO}=-I\ ,
\end{equation}
however, for a potential that averages over the
discontinuity~\cite{All2002MP}
\begin{equation}
  \label{smeq:ehomoav}
  \bar{\epsilon}_\mathrm{HOMO}=-I+\frac12\Delta_\mathrm{xc}\ .
\end{equation}
So, the DD is also given by
\begin{equation}
  \label{smeq:dxce}
  \Delta_\mathrm{xc}^\epsilon = 2(\bar{\epsilon}_\mathrm{HOMO} -
  \epsilon_\mathrm{HOMO})\ .
\end{equation}

In table~\ref{smtab:deltaxc} we give the values of the DD obtained with
the two previously mentioned approaches, Eq.~(\ref{smeq:dxc}) and
Eqs.~(\ref{smeq:ehomo})~and~(\ref{smeq:ehomoav}), for four different atoms.\\

\begin{table}
  \centering
  \begin{tabular*}{0.8\columnwidth}{@{\extracolsep{\fill}} ccccc}
    Atom & \(\Delta_\mathrm{xc}^\epsilon\) & \(\Delta_\mathrm{xc}^V\) \\
     \hline
     B &  0.281 & 0.284 \\
     C &  0.333 & 0.337 \\
     O &  0.427 & 0.435 \\
     F &  0.460 & 0.467 \\
   \end{tabular*}
   \caption{Comparison of methods to calculate the derivative
     discontinuity, Eq.~(\ref{smeq:dxc}) and Eq.~(\ref{smeq:dxce}).}
  \label{smtab:deltaxc}
\end{table}

\subsection{Computational cost of the correction}

In general the cost of the correction procedure is not significant in
comparison to the total cost of a DFT self-consistent procedure. The
additional cost comes from the numerical solution of
Eq.~(\ref{smeq:poisson_xc}). This is the same procedure required to
obtain the Hartree potential and can be done efficiently in linear or
quasi-linear computational time using methods like fast Fourier
transforms, multigrid or fast multipole expansions. Moreover, due to
linearity, it would be possible to solve the Poisson equation once,
obtaining at the same time the Hartree and the corrected XC
potentials.


\begin{table*}
\begin{tabular*}{0.95\textwidth}{@{\extracolsep{\fill}} lcccccccc}
\multirow{2}{*}{Molecule} & Experimental\(^a\) & \multicolumn{2}{c}{LDA} &
\multicolumn{5}{c}{CXD-LDA} \\
& \(E_g\)  & \(\Delta_\mathrm{ks}\) &	\(\Delta E_g\ \%\)
        & \(\eta_0\) & \(\Delta_\mathrm{ks}\) &\(\Delta_\mathrm{xc}\)	&
        \(E_g\) &	\(\Delta E_g\ \%\)\\
\cline{1-1}\cline{2-2}\cline{3-4}\cline{5-9}
H\(_2\)O	&	0.70	&	0.236	&	66	&	1.5E-03	&	0.249	&	0.356	&	0.605	&	13.6	\\
NH\(_3\)	&	0.60	&	0.204	&	66	&	6.4E-04	&	0.214	&	0.317	&	0.532	&	11.4	\\
CH\(_4\)	&	0.79	&	0.335	&	58	&	2.5E-04	&	0.348	&	0.294	&	0.642	&	18.8	\\
HF	&	0.81	&	0.322	&	60	&	4.1E-03	&	0.341	&	0.413	&	0.754	&	6.9	\\
PH\(_3\)	&	0.46	&	0.220	&	52	&	1.3E-04	&	0.225	&	0.256	&	0.482	&	4.7	\\
CO	&	0.58	&	0.245	&	58	&	3.8E-04	&	0.244	&	0.312	&	0.556	&	4.1	\\
N\(_2\)	&	0.65	&	0.286	&	56	&	3.7E-04	&	0.286	&	0.318	&	0.604	&	7.1	\\
H\(_2\)S	&	0.46	&	0.210	&	54	&	2.3E-04	&	0.210	&	0.274	&	0.484	&	5.2	\\
HCN	&	0.58	&	0.281	&	52	&	2.4E-04	&	0.282	&	0.292	&	0.574	&	1.1	\\
C\(_2\)H\(_2\)	&	0.52	&	0.245	&	53	&	1.7E-04	&	0.245	&	0.270	&	0.515	&	0.9	\\
C\(_2\)H\(_4\)	&	0.46	&	0.207	&	55	&	8.9E-05	&	0.208	&	0.255	&	0.463	&	0.6	\\
H\(_2\)CO	&	0.46	&	0.122	&	74	&	1.6E-04	&	0.121	&	0.287	&	0.408	&	11.3	\\
C\(_6\)H\(_6\)	&	0.38	&	0.188	&	51	&	1.3E-05	&	0.188	&	0.199	&	0.386	&	1.4	\\
\end{tabular*}
\caption{Experimental, LDA and CXD-LDA results for a set of
  molecules. 
  \(\eta_0\): optimized density threshold value from
  Eq.~(\ref{smeq:functional}), in all cases \(q_\mathrm{xc}(\eta_0)=-1\).
  \(\Delta_\mathrm{ks}\): HOMO-LUMO gap, \(\Delta_\mathrm{xc}\):
  derivative discontinuity, \(E_g\): fundamental gap, for CXD-LDA is
  \(\Delta_\mathrm{ks}+\Delta_\mathrm{xc}\), and \(\Delta E_g\): Difference in the gap with experimental results, for
  LDA \(\Delta_\mathrm{ks}\) is considered as the gap.
  All calculation are done with the Octopus code~\cite{Cas2006PSSB} using a grid composed of spheres of radius 16 a.u. around each atom with uniform
  spacing 0.25 a.u, and LDA norm-conserving
  pseudo-potentials. Geometries from Ref.~\cite{Cur1997JCP}. 
  \(^a\) Experimental values 
  compiled in Ref.~\cite{All2002MP}, except for C\(_6\)H\(_6\)
  obtained as difference between the ionization energy and the electron
  affinity from Refs.~\cite{NIST,Bur1987JCP}.}
\label{smtab:molecules}
\end{table*}

\begin{table*}
\begin{tabular*}{\textwidth}{@{\extracolsep{\fill}} lcccccccccccccc}
\multirow{2}{*}{Atom} & \multicolumn{2}{c}{Experimental\(^a\)}
&\multicolumn{4}{c}{LDA} &	\multicolumn{8}{c}{CXD-LDA}\\
&	\(I\)	&	\(E_g\)	&
\(\epsilon_\mathrm{homo}\)	&	\(\Delta I\ \%\)	&
\(\Delta_\mathrm{ks}\)	&	\(\Delta E_g\ \%\)	&
\(\eta_0\)	&	\(q_\mathrm{xc}(\eta_0)\)	&
\(\epsilon_\mathrm{homo}\)	&	\(\Delta I\ \%\)	&
\(\Delta_\mathrm{ks}\)	&	\(\Delta_\mathrm{xc}\)	&
\(E_g\)	&	\(\Delta E_g\ \%\)	\\
\cline{1-1}\cline{2-3}\cline{4-7}\cline{8-15}
H	&	0.500	&	0.472	&	-0.269	&	46	&	0.173	&	63	&	6.8E-04	&	-0.53	&	-0.421	&	15.8	&	0.174	&	0.307	&	0.481	&	2.0	\\
He	&	0.904	&	--	&	-0.570	&	37	&	0.570	&	--	&	8.2E-03	&	-0.63	&	-0.804	&	11.0	&	0.655	&	0.470	&	1.125	&	--	\\
Li	&	0.198	&	0.175	&	-0.116	&	41	&	0.042	&	76	&	1.1E-04	&	-0.69	&	-0.205	&	3.5	&	0.043	&	0.177	&	0.220	&	25.4	\\
Be	&	0.343	&	--	&	-0.206	&	40	&	0.129	&	--	&	7.1E-04	&	-0.82	&	-0.326	&	4.8	&	0.129	&	0.241	&	0.371	&	--	\\
B	&	0.305	&	0.295	&	-0.151	&	51	&	0.000	&	100	&	2.5E-03	&	-0.86	&	-0.291	&	4.5	&	0.000	&	0.284	&	0.284	&	3.5	\\
C	&	0.414	&	0.367	&	-0.227	&	45	&	0.000	&	100	&	6.2E-03	&	-0.92	&	-0.394	&	4.8	&	0.000	&	0.337	&	0.337	&	8.2	\\
N	&	0.534	&	0.534	&	-0.308	&	42	&	0.148	&	72	&	1.1E-02	&	-0.96	&	-0.502	&	6.1	&	0.148	&	0.391	&	0.539	&	0.9	\\
O	&	0.501	&	0.447	&	-0.272	&	46	&	0.000	&	100	&	2.2E-02	&	-1.00	&	-0.486	&	2.9	&	0.000	&	0.435	&	0.435	&	2.7	\\
F	&	0.640	&	0.515	&	-0.384	&	40	&	0.000	&	100	&	1.1E-02	&	-1.00	&	-0.614	&	4.1	&	0.000	&	0.467	&	0.467	&	9.4	\\
Ne	&	0.792	&	--	&	-0.498	&	37	&	0.498	&	--	&	1.0E-02	&	-1.00	&	-0.746	&	5.9	&	0.560	&	0.501	&	1.061	&	--	\\
Na	&	0.189	&	0.169	&	-0.113	&	40	&	0.034	&	80	&	1.1E-04	&	-0.69	&	-0.198	&	4.7	&	0.034	&	0.169	&	0.203	&	20.5	\\
Mg	&	0.281	&	--	&	-0.175	&	38	&	0.125	&	--	&	4.7E-04	&	-0.86	&	-0.282	&	0.5	&	0.126	&	0.215	&	0.341	&	--	\\
Al	&	0.220	&	0.204	&	-0.111	&	50	&	0.000	&	100	&	1.3E-03	&	-0.91	&	-0.222	&	0.8	&	0.000	&	0.224	&	0.224	&	9.8	\\
Si	&	0.300	&	0.249	&	-0.170	&	43	&	0.000	&	100	&	1.2E-03	&	-1.00	&	-0.297	&	1.0	&	0.000	&	0.256	&	0.256	&	2.9	\\
P	&	0.385	&	0.358	&	-0.231	&	40	&	0.079	&	78	&	6.8E-04	&	-1.00	&	-0.368	&	4.5	&	0.080	&	0.275	&	0.355	&	0.9	\\
S	&	0.381	&	0.304	&	-0.229	&	40	&	0.000	&	100	&	7.3E-04	&	-1.00	&	-0.375	&	1.6	&	0.000	&	0.293	&	0.293	&	3.6	\\
Cl	&	0.477	&	0.344	&	-0.305	&	36	&	0.000	&	100	&	7.3E-04	&	-1.00	&	-0.462	&	3.0	&	0.000	&	0.315	&	0.315	&	8.5	\\
Ar	&	0.579	&	--	&	-0.382	&	34	&	0.373	&	--	&	7.4E-04	&	-1.00	&	-0.549	&	5.2	&	0.394	&	0.335	&	0.729	&	--	\\
K	&	0.160	&	0.141	&	-0.096	&	40	&	0.023	&	84	&	7.5E-05	&	-0.74	&	-0.171	&	6.9	&	0.023	&	0.149	&	0.172	&	22.2	\\
Ca	&	0.225	&	0.224	&	-0.142	&	37	&	0.088	&	61	&	2.7E-04	&	-0.94	&	-0.232	&	3.4	&	0.089	&	0.182	&	0.271	&	21.0	\\
Sc	&	0.241	&	0.234	&	-0.141	&	42	&	0.000	&	100	&	4.0E-04	&	-0.95	&	-0.238	&	1.1	&	0.000	&	0.195	&	0.195	&	16.6	\\
Ti	&	0.251	&	0.248	&	-0.175	&	30	&	0.000	&	100	&	5.3E-04	&	-0.95	&	-0.277	&	10.2	&	0.000	&	0.205	&	0.205	&	17.3	\\
V	&	0.248	&	0.229	&	-0.186	&	25	&	0.000	&	100	&	6.4E-04	&	-0.94	&	-0.292	&	17.8	&	0.000	&	0.212	&	0.212	&	7.1	\\
Cr	&	0.249	&	0.224	&	-0.196	&	21	&	0.000	&	100	&	7.0E-04	&	-0.93	&	-0.305	&	22.7	&	0.000	&	0.219	&	0.219	&	2.3	\\
Mn	&	0.273	&	--	&	-0.206	&	25	&	0.045	&	--	&	8.4E-04	&	-0.92	&	-0.317	&	16.1	&	0.045	&	0.224	&	0.269	&	--	\\
Fe	&	0.290	&	0.285	&	-0.183	&	37	&	0.000	&	100	&	9.5E-04	&	-0.92	&	-0.299	&	2.9	&	0.000	&	0.233	&	0.233	&	18.2	\\
Co	&	0.290	&	0.265	&	-0.194	&	33	&	0.000	&	100	&	1.2E-03	&	-0.92	&	-0.314	&	8.5	&	0.000	&	0.241	&	0.241	&	9.2	\\
Ni	&	0.281	&	0.238	&	-0.205	&	27	&	0.000	&	100	&	1.3E-03	&	-0.92	&	-0.328	&	16.7	&	0.000	&	0.248	&	0.248	&	4.0	\\
Cu	&	0.284	&	0.239	&	-0.184	&	35	&	0.030	&	88	&	3.4E-03	&	-0.83	&	-0.309	&	8.8	&	0.030	&	0.253	&	0.283	&	18.5	\\
Zn	&	0.345	&	--	&	-0.223	&	35	&	0.176	&	--	&	1.7E-03	&	-0.91	&	-0.351	&	1.8	&	0.178	&	0.259	&	0.437	&	--	\\
Ga	&	0.220	&	0.205	&	-0.110	&	50	&	0.000	&	100	&	2.5E-03	&	-0.91	&	-0.226	&	2.3	&	0.000	&	0.236	&	0.236	&	15.2	\\
Ge	&	0.290	&	0.245	&	-0.165	&	43	&	0.000	&	100	&	1.6E-03	&	-1.00	&	-0.291	&	0.1	&	0.000	&	0.253	&	0.253	&	3.2	\\
As	&	0.361	&	0.331	&	-0.220	&	39	&	0.070	&	79	&	4.9E-04	&	-1.00	&	-0.351	&	2.7	&	0.070	&	0.262	&	0.332	&	0.5	\\
Se	&	0.358	&	0.284	&	-0.218	&	39	&	0.000	&		&	4.6E-04	&	-1.00	&	-0.355	&	1.0	&	0.000	&	0.272	&	0.272	&	4.3	\\
Br	&	0.434	&	0.311	&	-0.283	&	35	&	0.000	&	100	&	3.9E-04	&	-0.99	&	-0.425	&	2.1	&	0.000	&	0.285	&	0.285	&	8.2	\\
Kr	&	0.514	&	--	&	-0.346	&	33	&	0.332	&	--	&	3.6E-04	&	-1.00	&	-0.497	&	3.4	&	0.347	&	0.302	&	0.649	&	--	\\
Rb	&	0.154	&	0.136	&	-0.092	&	40	&	0.020	&	85	&	6.6E-05	&	-0.76	&	-0.164	&	6.7	&	0.020	&	0.144	&	0.164	&	21.0	\\
Sr	&	0.209	&	0.207	&	-0.132	&	37	&	0.070	&	66	&	2.5E-04	&	-0.96	&	-0.218	&	4.0	&	0.070	&	0.172	&	0.242	&	16.9	\\
\end{tabular*}
\caption{Experimental, LDA and CXD-LDA results for atomic
  systems. \(\eta_0\): optimized density threshold value from
  Eq.~(\ref{smeq:functional}), \(q_\mathrm{xc}(\eta_0)\): optimized
  value of the total XC charge [Eq.~(\ref{smeq:qxc})],
  \(\epsilon_\mathrm{HOMO}\): HOMO eigenvalue, \(I\):
  Ionization energy, \(\Delta I\): difference between
  \(-\epsilon_\mathrm{HOMO}\) and the experimental \(I\),
  \(\Delta_\mathrm{ks}\): HOMO-LUMO gap, \(\Delta_\mathrm{xc}\):
  derivative discontinuity, \(E_g\): fundamental gap, for CXD-LDA is
  \(\Delta_\mathrm{ks}+\Delta_\mathrm{xc}\), and \(\Delta E_g\): difference in the gap with experimental results, for
  LDA \(\Delta_\mathrm{ks}\) is considered as the gap.
  Values calculated using the APE code~\cite{Oli2008CPC} with spin-polarization.
  \(^a\) Experimental results from Ref.~\cite{NIST}, \(E_g\)
  values calculated as the difference of the ionization energy and the
  electron affinity.}
\label{smtab:atoms}
\end{table*}


\end{document}